# Virtual Critical Coupling in High-Power Resonant Systems

Aurora C. Araujo Martinez[1†*], Alex Krasnok[2,3†**], Sergey V. Kutsaev[1], Andrew Seltzman[4], Alexander Smirnov[1],

[1] *RadiaBeam, Santa Monica, CA 90404, USA*

[2] *Department of Electrical and Computer Engineering, Florida International University, Miami, FL 33174, USA*

[3] *Knight Foundation School of Computing and Information Sciences, Florida International University, Miami, FL 33199, USA*

[4] *MIT Plasma Science and Fusion Center, Cambridge, MA 02139, USA*

*†These authors contributed equally to this work.*

*To whom correspondence should be addressed:* * araujo@radiabeam.com, ** akrasnok@fiu.edu

## Abstract

Exciting high-power resonators pose challenges such as managing power reflections, which can cause energy losses and damage system components. This is crucial for applications like Lower Hybrid Current Drive (LHCD) systems in tokamaks, where plasma stability and confinement depend on efficient energy transfer. In this work, we introduce the Virtual Critical Coupling (VCC) mechanism to address reflection-related challenges in S-band resonators. We theoretically designed a complex frequency excitation signal tailored to the resonator's characteristics, facilitating efficient energy storage and minimizing reflections without mechanical modifications. Using a custom low-level RF system, we conducted experiments at 32 mW and 600 kW with a 5 MW S-band klystron, demonstrating a ninefold reduction in reflection coefficients compared to traditional monochromatic excitation in high-power tests. This approach enhances the efficiency and stability of high-power resonant systems, potentially advancing nuclear fusion energy production.

## Introduction

Efficient, reflection-free excitation of high-power resonators and resonant systems is crucial in various fields, including high-power microwave systems, particle accelerators, high-frequency wireless transceivers, and nuclear fusion [1–6]. In high-power microwave systems, reducing reflections prevents energy losses and potential component damage [7–11]. Similarly, in particle accelerators, minimizing reflections ensures effective energy transfer to accelerating structures, optimizing performance [5,6]. Extending these principles to the field of nuclear fusion, particularly within tokamak reactors, Lower Hybrid Current Drive (LHCD) and Ion Cyclotron Radio Frequency (ICRF) systems underscores the necessity for efficient excitation fed through coaxial transmission lines or waveguides, respectively. Noteworthy achievements of the LHCD system include driving the largest non-inductive current of 3.6 MA in JT-60U, sustaining current for up to two hours in TRIAM-1M, and attaining the highest current drive efficiency to date [12,13]. These advancements highlight that ensuring reflection-free excitation of high-power microwave fields is paramount for LHCD in tokamak reactors, as it guarantees maximal power transfer to the plasma, optimizes current drive efficiency, and maintains plasma stability. ICRF will be required for ignition in the upcoming SPARC tokamak [14,15] and future tokamak power reactors.

Despite these successes, operating below the plasma edge density cutoff introduces significant sensitivity to density fluctuations, which can increase return loss and elevate peak electric fields within the multi-junction structures beyond the vacuum multipactor threshold of ~9.3kV/cm at 4.6 GHz [16–18] limiting the maximum power handling capability of the launching structure. This sensitivity poses a critical challenge for maintaining consistent performance in LHCD systems. To address this issue, RadiaBeam in collaboration with the MIT Plasma Science and Fusion Center are planning to develop an improved High Field Side (HFS) LHCD launcher module for the DIII-D tokamak [19]. This initiative leverages the advantages of HFS launch, such as enhanced wave coupling and increased current drive efficiency, thereby advancing plasma performance optimization in fusion reactors. The current DIII-D LHCD system [20] is engineered to couple 4.6 GHz microwaves into an optimized spectrum for achieving non-inductive off-axis current generation and comprises eight multijunction toroidally adjacent modules, each capable of handling up to 200 kW.

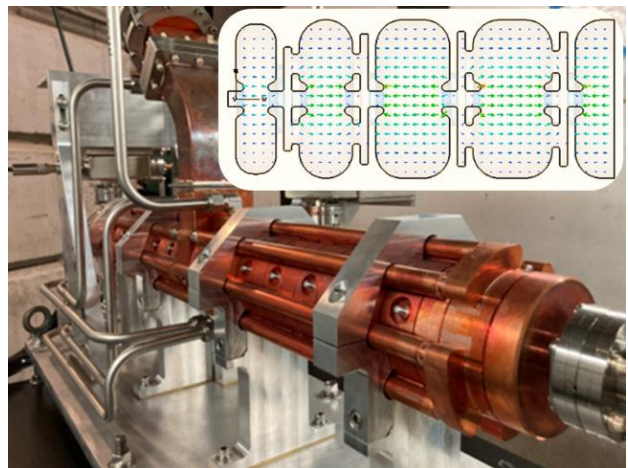

**Figure 1.** RadiaBeam's S-band coupled-cavity resonator. Inset: electric field distribution inside its first four cavities.

To overcome the challenges associated with reflection and energy loss in high-power microwave excitation systems, we introduce the Virtual Critical Coupling (VCC) mechanism. Building upon the concept of virtual absorption [21,22], VCC facilitates transient energy transfer by temporally shaping the excitation signal, effectively eliminating reflection and transmission losses. Unlike traditional critical coupling techniques that rely on steady-state conditions, VCC exploits transient phenomena during the energy coupling process, allowing incident energy to be fully absorbed and stored within the resonator during the excitation pulse. This is achieved through the temporal modulation of the excitation signal with a precise complex frequency characterized by an exponential growth in amplitude. Such modulation balances the leakage of signal energy through destructive interference, enabling excitation without scattering [10,11,23,24]. This delicate balance is attainable at specific complex frequencies corresponding to the zeros of the scattering matrix (S-matrix), which are intrinsically linked to the system's singularities [21].

In this work, we apply the VCC mechanism to achieve reflection-free excitation in high-power S-band resonators. Our primary objective is to temporally shape a complex frequency excitation signal to ensure that all impinging energy is trapped without reflection. To accomplish this, we developed a tailored RF circuit capable of generating optimized excitation signals and conducted comprehensive experimental demonstrations using a high-power 5 MW S-band klystron. We began by optimizing an exponential excitation signal based on Temporal Coupled Mode Theory [10,25,26], which accounts for the resonators' characteristic frequency, coupling parameters, and quality factors (Q-factors). Subsequently, we designed, built, and tested a low-level radio frequency (LLRF) circuit [27] to generate the VCC pulse signal, ensuring its functionality across both low and high-power scenarios. Comprehensive experimental demonstrations were conducted using a high-power 5 MW S-band klystron. At low power, the LLRF circuit successfully reduced the reflection coefficient by over 70% compared to monochromatic excitation signals. At high power (600 kW), the VCC mechanism achieved a ninefold reduction in reflection coefficients, validating its efficacy in minimizing power reflections and enhancing the performance of high-power resonant systems. By demonstrating the VCC mechanism in an S-band high-power (600 kW level) resonator, our work represents a significant step toward enhancing the efficiency of both particle accelerators and LHCD systems.

## Results

Microwave resonators and RF sources are essential in accelerator and fusion technologies for efficient energy transfer and precise electromagnetic field control. In accelerators, resonant cavities transfer RF power to charged particle beams for acceleration [28]. In fusion reactors, antennas couple RF energy into plasma for heating and current drive [29]. Both applications utilize high-power RF sources, such as klystrons or solid-state amplifiers, to maintain precise resonant frequencies. In this work, we utilized the accelerator technology resources and facilities at RadiaBeam to experimentally demonstrate the virtual critical coupling (VCC) mechanism at power levels compatible with Lower Hybrid Current Drive (LHCD) technology. The resonator employed consists of multiple pillbox cavities electromagnetically coupled to function collectively as a single resonant structure capable of supporting various excitation modes, Figure 1. Each mode corresponds to a distinct phase difference between the individual cavities, such as 0 or π radians, allowing selective excitation of specific resonant states. The specific resonator utilized in our experiments is characterized by the following microwave parameters: a resonant frequency of 2856 MHz for the π/2-mode, a unloaded quality factor ($Q_0$) of 13,600, and a single input port for power feeding with a coupling coefficient ($\beta$) of 2.43. Such a high Q-factor indicates a low rate of energy loss relative to the stored energy, reflecting the resonator's ability to maintain oscillations over time. A coupling coefficient of $\beta = 2.43$ signifies that the resonator is over-coupled, meaning that the power input into the resonator exceeds the power dissipated intrinsically, which is advantageous for efficient energy transfer in high-power applications. This resonator, originally designed for particle acceleration, is compatible with the frequency band and power requirements of certain LHCD systems, such as the 2.45 GHz LHCD antenna on EAST tokamak [30], making it a suitable candidate for the experimental validation of the VCC mechanism. The ability to support multiple excitation modes through phase differences between coupled cavities not only enhances the flexibility of the resonator but also ensures that the VCC mechanism can be effectively tested under conditions that closely mimic those encountered in practical applications.

We begin our analysis with a theoretical investigation using Temporal Coupled-Mode Theory (TCMT), a robust and intuitively transparent framework for examining the dynamic behavior of resonant systems interacting with external fields [10,26,31,32]. TCMT leverages experimentally measurable parameters, such as transmission and reflection coefficients, Q-factors, and coupling coefficients, to accurately describe specific experimental setups. This theory not only accommodates linear interactions but also can be extended to account for nonlinear effects arising from plasma or resonator changes. Moreover, TCMT's versatility allows it to handle excitation pulses of arbitrary shapes, including complex frequency signals, thereby providing the necessary flexibility for implementing VCC. To accurately model the dynamic behavior of our test cavity at the RadiaBeam facility within the TCMT framework, we first incorporate the cavity's specific parameters. The resonant frequency of the cavity is $f_0 = 2856$ MHz, and it operates with an unloaded quality factor of $Q_0 = 13,600$. Additionally, the cavity exhibits a waveguide coupling coefficient of $\beta = 2.43$, indicating an over-coupled regime. The unloaded quality factor $Q_0$ is related to the intrinsic decay rate $\gamma_i$ by the equation $Q_0 = \omega_0 / 2\gamma_i$, where $\omega_0 = 2\pi f_0$ is the angular resonant frequency. Solving for $\gamma_i$, we obtain the intrinsic decay time, $\tau_i = 1/\gamma_i = 2Q_0/\omega_0 = 4,329T$, with $T = 1/f_0$ representing the period of the resonant frequency. The waveguide coupling coefficient $\beta$ can be defined as the ratio of the unloaded quality factor to the external quality factor $Q_e$, such that $\beta = Q_0/Q_e$. Solving for $Q_e$, we obtain $Q_e = Q_0/\beta$, which leads to an external decay time $\tau_e = 1/\gamma_e = 2Q_e/\omega_0 = 1,780T$. The coupling rate $\rho$ is then calculated using the known TCMT relation $\rho = \sqrt{2/\tau_e} = \sqrt{2\gamma_e}$ [32].

In the TCMT framework, the resonator is coupled to an external port through which an incoming electromagnetic field $S_{in}(t)$ is introduced. This port facilitates the exchange of energy between the resonator and the external environment. The resonator experiences both external $\gamma_e$ and intrinsic $\gamma_i$ loss rates. The dynamic behavior of the resonator can be described by the differential equation for the amplitude $a(t)$ of the resonant mode:

$$\frac{da(t)}{dt} = \left(-i\omega_0 - \gamma_e - \gamma_i\right)a(t) + \sqrt{2\gamma_e}\,S_{in}(t) \tag{1}$$

The term $-i\omega_0 a(t)$ represents the natural oscillation of the resonator at its resonant frequency $\omega_0$, the decay terms $-\gamma_e a(t)$ and $-\gamma_i a(t)$ account for energy dissipation through external coupling and intrinsic mechanisms, respectively, the term $\sqrt{2\gamma_e}\,S_{in}(t)$ models the excitation of the resonator by the incoming field $S_{in}(t)$. The relationship between the incoming field $S_{in}(t)$ and the reflected field $S_{ref}(t)$ is given by:

$$S_{ref}(t) = -S_{in}(t) + \sqrt{2\gamma_e}\,a(t) \tag{2}$$

The reflected field consists of the unperturbed incoming field minus the resonator's field scattered back through the port. The objective is to design the input signal $S_{in}(t)$ such that all the impinging energy is absorbed by the resonator without any reflection, i.e., $S_{ref}(t) = 0$ in a steady state.

We investigated $S_{ref}(t)$ when the cavity was subjected to two distinct types of excitation pulses: a monochromatic continuous wave (CW) pulse and a complex frequency excitation pulse tailored to achieve VCC. The monochromatic CW pulse is defined as $S_m(t) = e^{i\omega t}\cdot \text{UnitStep}(15000T - t)$, where $\omega$ is the angular frequency of the excitation, and $\text{UnitStep}(15000T - t)$ represents a unit step function that activates the pulse at $t = 0$ and deactivates it at $t = 15000T = 3.5\,\mu s$. This signal corresponds to the standard step charging of resonators [6]. In contrast, the complex frequency excitation pulse is formulated as $S_{ex}(t) = S_0 \cdot e^{i\omega t} \cdot e^{t(\gamma_e - \gamma_i)} \cdot \text{UnitStep}(15000T - t)$, where $S_0$ is the initial amplitude of the exponential growth, introduced to initiate the signal from a non-zero value. The exponential term $e^{t(\gamma_e - \gamma_i)}$ is engineered to match the resonator's complex zero, effectively shaping the excitation signal to compensate for both external and intrinsic decay rates of the cavity. The primary objective of employing the complex frequency excitation pulse $S_{ex}(t)$ is to achieve VCC by ensuring that the energy impinging on the resonator is fully absorbed without reflection [10].

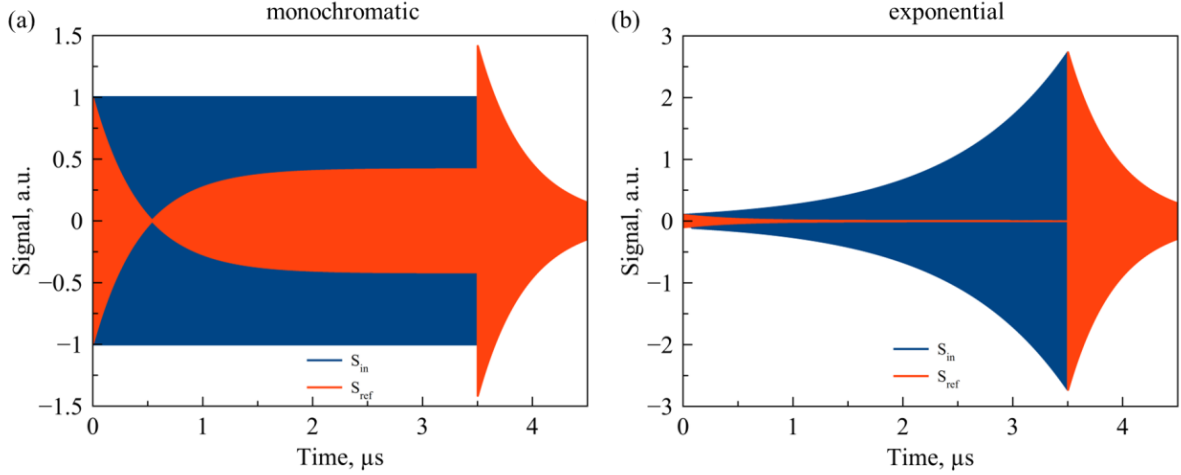

**Figure 2**. Numerical simulation of reflected signals for different excitation scenarios. (a) Reflection response (red) to a monochromatic CW pulse (blue), showing an instantaneous matching time $t_{r0} = 0.54\,\mu s$ and a steady-state reflection level $r_\infty = 0.4$. (b) Reflection response (red) to a complex frequency excitation pulse (blue) designed for VCC, illustrating an initial nonzero reflection that rapidly decays to zero, thereby confirming effective suppression of reflections.

We utilized Mathematica's `NDSolve` function to numerically solve Eqs. (1) and (2) for the monochromatic and exponential incoming excitation signals. The reflection response to the monochromatic pulse is presented in Figure 2(a). When the source initiates the excitation of fields within the resonator, a portion of the supplied power is reflected and dissipated. This occurs because an unloaded cavity behaves as a mismatched load, leading to initial reflections. This phenomenon is particularly significant in high-Q cavities, which possess extremely high-quality factors. As the electromagnetic fields gradually build up within the resonator, the reflection first diminishes at distinct instantaneous matching time of $t_{r0}$ (0.54 μs in our case) and then grows again to the steady-state reflection level denoted by $r_\infty$ (0.4 in our case). These parameters are crucial for experimentally determining the necessary adjustments to the complex frequency excitation pulse [11]. The instantaneous matching occurs when the resonator is sufficiently excited such that its radiated field destructively interferes with the directly reflected wave, i.e., $\sqrt{2\gamma_e}\,a(t_{r0}) = S_m(t_{r0})$, where $a(t_{r0})$ is the resonator amplitude at the instantaneous matching time. Upon the abrupt cessation of the excitation wave, the energy stored within the resonator begins to dissipate, resulting in an exponentially decaying release of energy. Zero reflection in the steady-state regime can be achieved by extending the instantaneous matching time to infinity, corresponding to the conventional critical coupling regime ($\gamma_e = \gamma_i$). However, this approach does not result in more efficient resonator excitation because it either introduces losses within the resonator or leads to stronger reflections during the mode buildup phase [10].

The response of the resonator to the complex frequency excitation pulse, as depicted in Figure 2(b), highlights the effectiveness of the VCC mechanism. Upon the initiation of the incoming complex frequency pulse, a nonzero reflection is initially observed. This initial reflection arises because the excitation starts from a finite small amplitude $S_0$, which does not immediately satisfy the VCC condition. However, by initiating the excitation pulse with a smaller amplitude $S_0$, the reflected signal can be minimized from the outset. As the excitation signal undergoes exponential growth, the initially weak reflection rapidly diminishes. As a result, the total reflection attains a steady-state value of zero, confirming that the excitation has successfully engaged the complex zero of the resonator. This complete absorption of the impinging energy without reflection is a definitive indication of the VCC condition being met. Upon cessation of the excitation pulse, the energy stored within the resonator begins to be released.

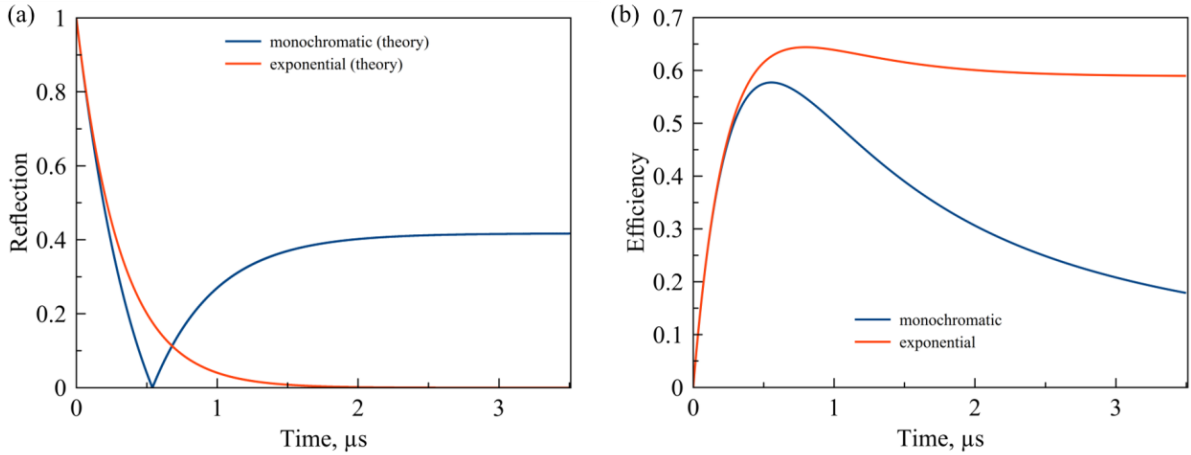

**Figure 3**. (a) Reflection coefficient (reflected/incoming signal ratio) for monochromatic and exponential excitation pulses. (b) Excitation efficiency over time, reaching approximately 20% for monochromatic excitation and 60% for exponential excitation in a lossy resonator.

Figure 3(a) presents the reflection coefficient, defined as the ratio of reflected to incoming signals, for both monochromatic and exponential excitation scenarios. The results clearly illustrate that, although the reflection coefficient initially decays more rapidly under monochromatic excitation, the exponential excitation achieves a substantially lower reflection level immediately following the instantaneous matching effect. Specifically, while the monochromatic case exhibits a persistent steady-state reflection, the exponential excitation effectively suppresses reflections, approaching zero. The excitation efficiency $\eta(t)$, defined as the ratio of the instantaneous stored energy within the resonator $|a(t)|^2$ to the total energy delivered to the resonator $\int_0^t |S_{in}(t')|^2 dt'$:

$$\eta(t) = \frac{|a(t)|^2}{\int_0^t |S_{in}(t')|^2 dt'} \quad (3)$$

serves as a critical metric for evaluating the effectiveness of resonator excitation. In an ideal scenario, where all input energy is perfectly absorbed by the resonator, the excitation efficiency attains a value of unity. Recent studies [10] have demonstrated that achieving this perfect efficiency is feasible exclusively through the implementation of VCC in conjunction with a lossless resonator. Figure 3(b) illustrates the calculated excitation efficiency as a function of time for both monochromatic and exponential excitation pulses applied to our realistically lossy resonator. Under monochromatic excitation, the efficiency initially increases as energy accumulates within the resonator. However, upon reaching steady-state conditions, the efficiency stabilizes at approximately 20%. This plateau indicates the presence of inherent losses within the resonator, which limit the maximum achievable efficiency. In contrast, when employing an exponential excitation pulse designed to fulfill the VCC conditions, the excitation efficiency exhibits a markedly superior performance, reaching up to 60%. This substantial enhancement is attributable to the tailored temporal shaping of the excitation signal, which effectively compensates for both external and intrinsic decay rates of the resonator and minimizing the transient reflection inherent to the standard step charging of resonators [6].

To experimentally validate the VCC regime, we developed a low-level radio frequency (LLRF) system tailored to generate the VCC pulse signal for the 2856 MHz test resonator. The LLRF circuit was engineered to enable initial verification of the VCC regime at low power levels, utilizing amplifiers to produce signals in the milliwatt range. This design allowed for seamless transition to high-power testing by employing a 5 MW klystron with the same circuit, thereby demonstrating the scalability and robustness of the VCC implementation. The VCC experiments were conducted using a suite of precision instruments, including a GW INSTEK AFG-2225 Arbitrary Function Generator and an AFG Wavetek 50 MHz Model 801 Pulse Generator to create the necessary excitation signals. An Anritsu MG3692B 20 GHz Signal Generator served as the local oscillator (LO) source, while an Agilent Infinitum DSO81204B 12 GHz 40 GSa/s oscilloscope was employed to accurately measure both the forward and reflected signals. More details of the experimental setup are provided in the Supplementary Materials.

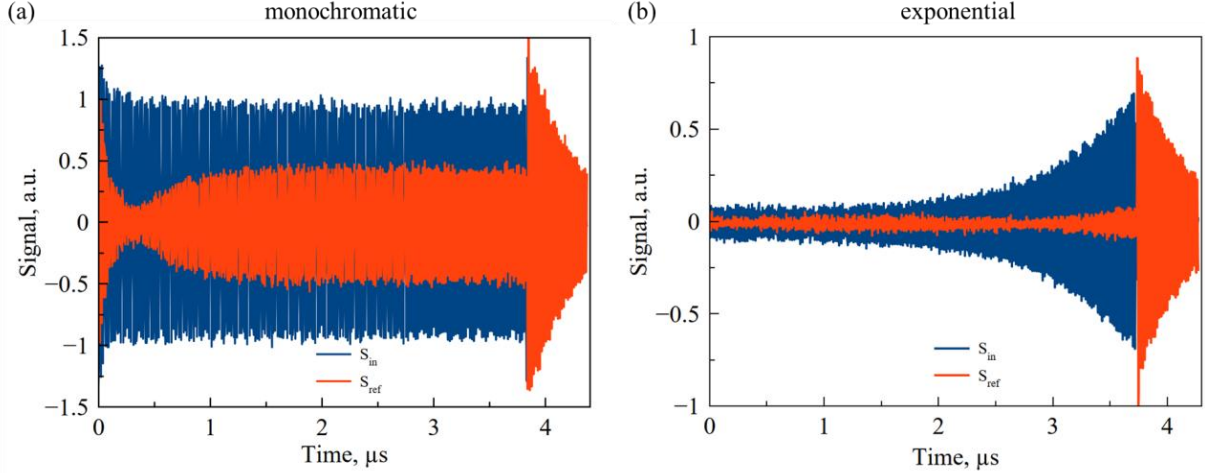

**Figure 4.** Low-power (~32 mW) experimental time-domain responses of the resonator to (a) monochromatic CW excitation and (b) complex frequency excitation designed for VCC.

To experimentally validate the VCC regime, we first commenced with low-power tests. The resonator was driven by a monochromatic excitation signal, and both the incident ($S_{in}$) and reflected ($S_{ref}$) signals are shown in Figure 4(a). Specifically, a step pulse with constant amplitude was generated using an AFG and applied to the resonator. The Mini-Circuits Amplifiers (ZRL-3500+) were used to provide ~32 mW (15 dBm) power. The resulting $S_{in}$ and $S_{ref}$ signals were captured using a high-speed oscilloscope. The amplitude of the signal was adjusted so that the measured signals in the scope were below its limit of 5V. The experimental results demonstrate strong concordance with the theoretical predictions presented in Figure 2(a), notably featuring an instantaneous matching time of $t_{r0} = 0.5$ μs and a steady-state reflection level of $r_\infty = 0.4$.

These experimentally determined parameters are pivotal for defining the complex frequency required to achieve VCC through complex zero excitation [11,23]. Specifically, using the measured values $r_\infty = 0.4$ and $t_{r0} = 0.54$ μs, we calculated the internal decay rate $\gamma_{int}$ and the external decay rate $\gamma_{ext}$ using $\gamma_{int} = \frac{1-r_\infty}{2t_{r0}}\ln\left(\frac{1+r_\infty}{r_\infty}\right)$ and $\gamma_{ext} = \frac{1+r_\infty}{2t_{r0}}\ln\left(\frac{1+r_\infty}{r_\infty}\right)$. Substituting the experimental values yields: $f_{zero} = f_0 + i(\gamma_{ext} - \gamma_{int}) \approx 2.856 \times 10^9 + i9.4 \times 10^5$ rad/s. This calculated $f_{zero}$ is consistent with the theoretical prediction based on the relation, $f_{zero} = 2.856 \times 10^9 + i(\gamma_e - \gamma_i) = 2.856 \times 10^9 + i9.45 \times 10^5$ rad/s.

To achieve the VCC effect at low power, we utilized the determined complex frequency by loading the optimized exponential signal envelope into the AFG. This approach enabled the precise amplitude modulation necessary for VCC, and the resulting signals are depicted in Figure 4(b). The experimental outcomes closely mirror the theoretical predictions presented in Figure 2(b), demonstrating the effective implementation of the VCC mechanism.

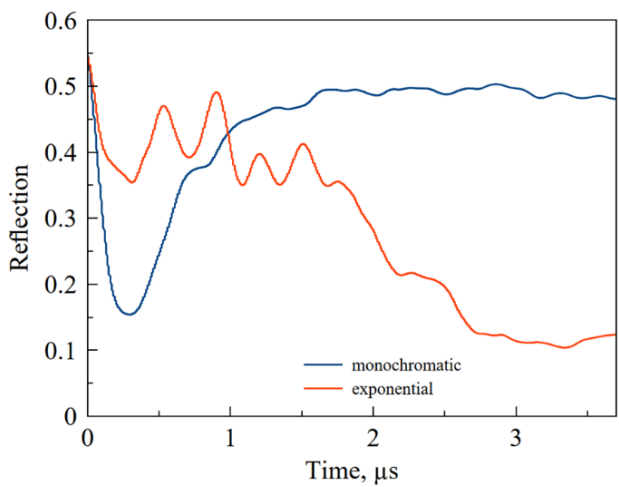

**Figure 5.** Experimental reflection coefficients for monochromatic and exponential excitation pulses, illustrating a significant reduction in reflections achieved through the VCC mechanism.

To accurately determine the reflection coefficient from the measured signals, we developed a Python script for post-processing. The script first computed the analytic signal using the Hilbert transform. The envelope of the signal was then extracted as the magnitude of the analytic signal. To enhance the quality of the envelope by reducing noise, a Savitzky-Golay filter was applied with a specified window length and polynomial order. This filtering process produced a smooth and refined envelope suitable for subsequent analysis. Using the processed envelopes, we calculated the reflection coefficients for both exponential and monochromatic excitations, Figure 5. The reflection coefficient under exponential excitation exhibited a significant reduction (70%) compared to the monochromatic case, corroborating the efficacy of the VCC mechanism in minimizing power reflections.

Having successfully demonstrated the VCC regime at low power, we proceeded to evaluate the LLRF circuit under high-power conditions. The high-power VCC test was conducted with the resonator installed in one of RadiaBeam's bunkers and conditioned to operate under vacuum at elevated power levels. Adhering to the same methodology employed in the low-power experiments, we ensured consistency in our testing approach. Detailed schematics of the experimental setup for the high-power test, including the 5 MW ScandiNova K1 klystron amplifier and the linear accelerator (linac), are provided in the Supplementary Materials.

During the high-power VCC test, the input power was ramped up to 600 kW, delivered in 4 μs pulses at a repetition rate of 1 Hz. To accurately capture the forward and reflected signals using the high-speed oscilloscope, two -80 dB attenuators were integrated into the outputs of the bidirectional coupler. This attenuation was necessary to ensure that the measured signal amplitudes remained within the dynamic range of the oscilloscope, thereby facilitating precise signal acquisition despite the high-power levels involved.

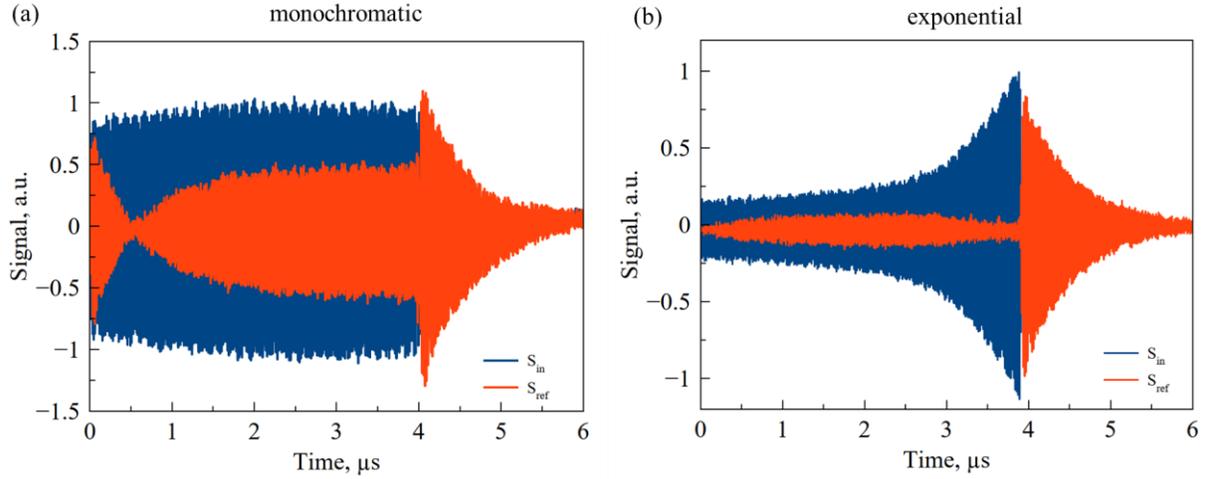

**Figure 6.** Time-domain responses of the resonator under high-power (600 kW) conditions to (a) monochromatic CW excitation and (b) complex frequency excitation designed for VCC, demonstrating effective suppression of reflections in the high-power VCC regime.

The experimental results of the high-power tests are presented in Figure 6, which illustrates the time-domain responses of the resonator to both monochromatic CW excitation and complex frequency excitation designed for VCC. Figure 6(a) shows the reflection response to the monochromatic CW pulse, where an instantaneous matching time and a steady-state reflection level are observable, consistent with the low-power and theoretical results. In contrast, Figure 6(b) depicts the reflection response under complex frequency excitation, demonstrating the effective suppression of reflections as predicted by the VCC mechanism. The high-power experiments corroborate the low-power findings, validating the scalability and robustness of the VCC approach in mitigating power reflections in high-power resonant systems.

Figure 7 presents the reflection coefficients for both exponential and monochromatic excitation pulses during the high-power experiments. The system reaches a steady state at approximately 2 μs, at which point reflections begin to decrease for the exponential excitation. By 3.5 μs, the exponential excitation achieves the VCC regime. Notably, the optimized VCC exponential excitation signal reduces reflections by more than ninefold compared to the typical monochromatic excitation at steady state. These results conclusively demonstrate the successful implementation of the VCC regime under high-power conditions, achieving effective reflection suppression with input powers up to 600 kW.

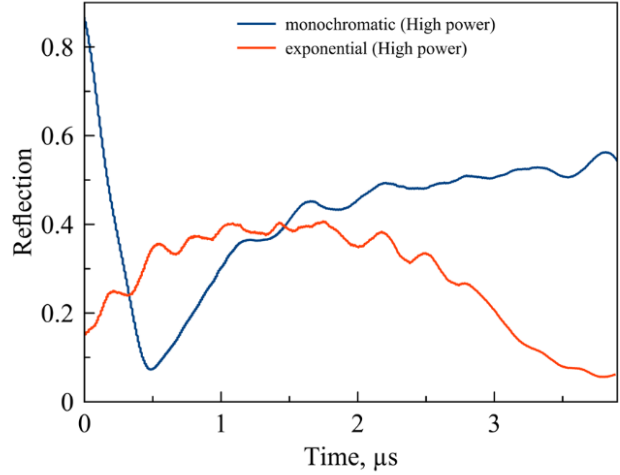

**Figure 7.** Reflection coefficients for exponential and monochromatic excitation pulses during high-power tests. Exponential excitation achieves virtual critical coupling, reducing reflections by over ninefold compared to monochromatic excitation at steady state with input powers up to 600 kW.

## Discussion

The application of the VCC mechanism presents a promising approach to optimizing high-power resonant systems used in accelerator technologies and fusion reactors. By achieving conditions with minimal reflections, VCC facilitates improved energy transfer efficiency and reduces the potential for performance issues and component wear due to reflected power. This is particularly relevant for LHCD systems, where precise current drive and plasma stability are important for maintaining sustained fusion reactions. Compared to traditional critical coupling techniques, which often require complex mechanical adjustments and are sensitive to varying plasma surface impedance under changing plasma density conditions [33], VCC offers a more adaptable solution through temporal signal shaping. This approach allows for maintaining optimal coupling despite environmental and operational changes, enhancing the system's ability to handle transient disturbances commonly encountered in tokamak operations. A resonant structure that matches each aperture to the plasma load reduced circulating power within the vacuum section of the multijunction. The standing-wave induced by this circulating power increases the peak electric field within the vacuum section of the multijunction placing limits in the maximum power handling capacity limited by multipactor breakdown. Time-domain variation of incident power to a multijunction is an under-explored area of LHCD physics as the majority of experiments utilize an un-modulated CW signal of constant frequency; exploration of time-domain modulation of LHCD drive power enables two electrically tunable variables influencing coupling efficiency that permit lower circulating power and improve system reliability.

Nevertheless, the implementation of VCC involves certain challenges. Achieving precise control over the excitation signal's complex frequency is crucial for ensuring the desired destructive interference and energy absorption. This requires advanced signal generation and real-time monitoring capabilities, which may add complexity and cost to the system. Additionally, while the experiments conducted showed substantial reflection suppression, further research is needed to assess the long-term stability and reliability of VCC under continuous high-power operation and varying plasma conditions. Future work should explore the application of VCC across a wider range of resonator types and frequency bands to broaden its applicability in different high-power systems. Integrating adaptive techniques, such as machine learning algorithms for signal shaping [34–36], could enhance the responsiveness and precision of VCC in real-time environments. Additionally, investigating the combination of VCC with other current drive techniques may provide complementary benefits, potentially improving the overall performance of fusion and accelerator technologies.

## Conclusion

In this work, we have demonstrated the effectiveness of the VCC mechanism in reducing power reflections within high-power S-band resonators. By utilizing a carefully shaped complex frequency excitation signal, the VCC mechanism was able to suppress reflections by over ninefold compared to traditional monochromatic excitation in both low-power (32 mW) and high-power (600 kW) experimental setups. The development and validation of an LLRF system further highlight the practicality and scalability of VCC for integration into existing high-power resonant systems without requiring mechanical modifications. These findings contribute to enhancing the performance and reliability of magnetic confinement devices, supporting the advancement of more efficient and stable plasma operations essential for nuclear fusion energy production.

# Supplementary Materials:

# Virtual Critical Coupling in High-Power Resonant Systems

Aurora C. Araujo Martinez[1†*], Alex Krasnok[2,3†**], Sergey V. Kutsaev[1], Andrew Seltzman[4], Alexander Smirnov[1],

[1] *RadiaBeam, Santa Monica, CA 90404, USA*

[2] *Department of Electrical and Computer Engineering, Florida International University, Miami, FL 33174, USA*

[3] *Knight Foundation School of Computing and Information Sciences, Florida International University, Miami, FL 33199, USA*

[4] *MIT Plasma Science and Fusion Center, Cambridge, MA 02139, USA*

*†These authors contributed equally to this work.*

*To whom correspondence should be addressed:* * araujo@radiabeam.com, ** akrasnok@fiu.edu

**Experimental Setup**

To achieve the exponential amplitude modulation necessary for Virtual Critical Coupling (VCC), we initially employed a voltage variable dual control attenuator (VVDCA). This specialized electronic component adjusts the attenuation of two independent signal paths based on a control voltage. The corresponding low-level radio frequency (LLRF) circuit is depicted in **Figure S1**. It consists of a local oscillator (LO) generating a continuous wave (CW) sinusoidal signal at 2856 MHz, followed by a filter, a switch connected to a pulse generator to create a 5 µs pulse, and an amplifier. A circulator with a load prevents reflected power from damaging the amplifier. The VVDCA modulates the amplitude of the signal, while two directional couplers monitor the input forward (FWD) and reflected (REF) signals. The final component in this circuit is the 2856 MHz resonator cavity, which operates within the S-band frequency range. This approach aimed to validate the theoretical predictions outlined in the main text regarding transient energy absorption and reflection minimization. An arbitrary function generator (AFG) with two channels was used to control the VVDCA: one channel generated an exponential voltage function, while the other provided a rectangular pulse with constant voltage to regulate attenuation.

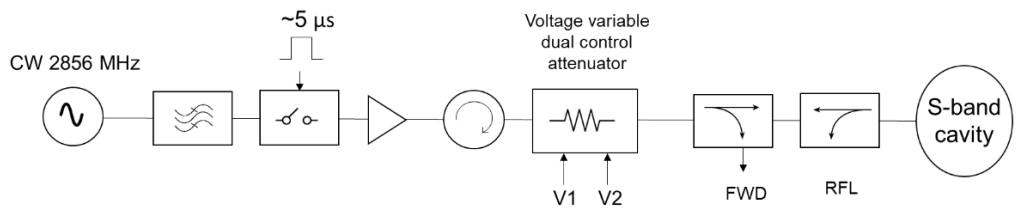

**Figure S1.** LLRF circuit using a voltage variable dual control attenuator. This circuit was designed to modulate the input signal amplitude for VCC experiments.

**Figure S2** presents the VVDCA-based LLRF circuit built at RadiaBeam. While this configuration successfully generated an exponentially increasing signal, precise amplitude control to achieve optimal VCC remained challenging. To address this, we modified the circuit to implement in-phase and quadrature (I/Q) modulation, which allows independent control of amplitude and phase modulation.

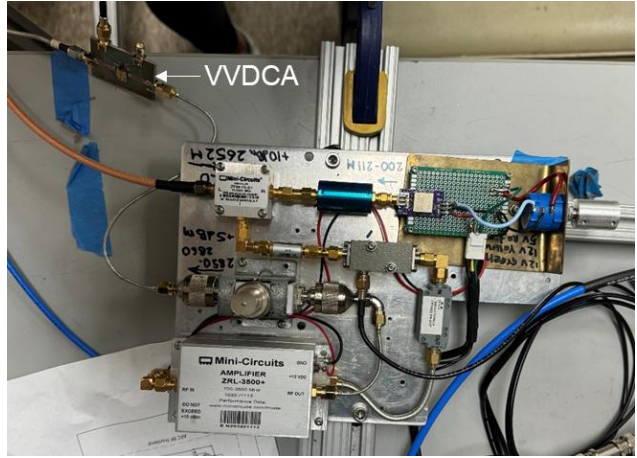

**Figure S2.** LLRF circuit using a voltage variable dual control attenuator built by RadiaBeam. This circuit enabled initial amplitude modulation experiments but required further refinement for optimal VCC implementation.

**Figure S3** illustrates the revised LLRF system incorporating I/Q modulation. A pulse generator produced a 5 μs pulse, while an AFG generated an optimized exponential envelope tailored to the test cavity's parameters to achieve VCC. This envelope was created using a Python script and loaded into the AFG. The signal was then mixed with an LO carrier wave at 2856 MHz, filtered, and amplified before being transmitted to the cavity. A circulator with a load protected the amplifier, and directional couplers monitored the FWD and REF signals using a high-speed oscilloscope. For low-power testing, Mini-Circuits amplifiers (ZRL-3500+) were used, as shown in **Figure S4**. The introduction of I/Q modulation was crucial in ensuring that the excitation signal precisely matched the theoretical complex frequency predicted to engage the VCC condition. This enhancement directly correlates with the improved efficiency and reflection suppression discussed in the main text. For high-power testing, the circuit was adapted to handle increased power levels by replacing the amplifier, circulator, load, and couplers with components rated for 5 MW operation.

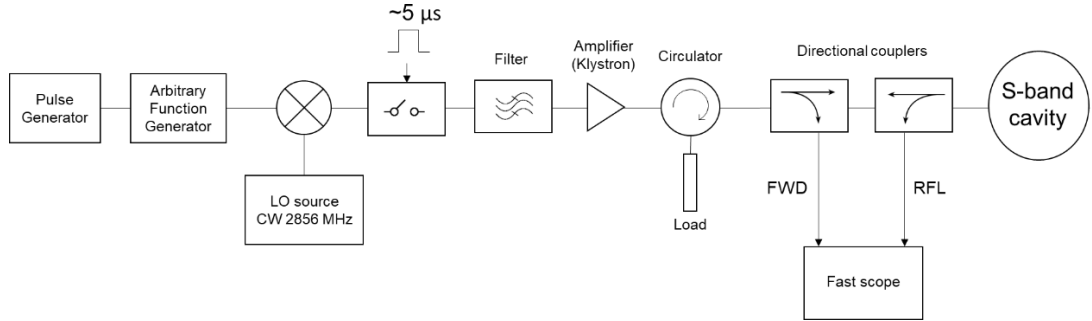

**Figure S3.** LLRF system for VCC driver signal. This system implemented I/Q modulation to precisely shape the excitation signal for optimal VCC conditions.

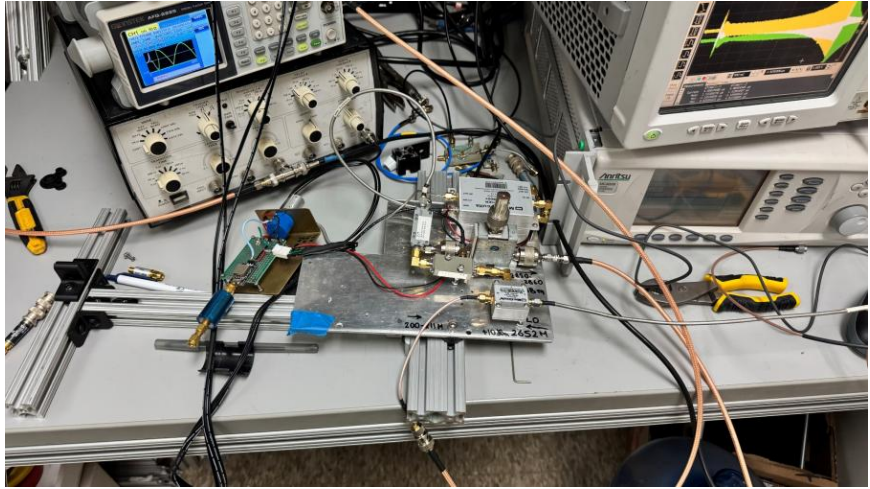

**Figure S4.** LLRF circuit used for low-power VCC test. The circuit employed Mini-Circuits amplifiers and directional couplers to enable initial validation of the VCC mechanism.

**Low-Power Experimental Validation**

To validate the VCC regime, we conducted experiments using the following equipment:

- **Arbitrary Function Generator:** GW INSTEK AFG-2225

- **Pulse Generator:** AFG Wavetek 50 MHz Model 801

- **LO Source:** Anritsu MG3692B 20 GHz Signal Generator

- **High-Speed Oscilloscope:** Agilent Infinitum DSO81204B (12 GHz, 40 GSa/s)

Before performing the experiment, we simulated the expected FWD and REF signals using the VCC model. The optimized exponential envelope was loaded into the AFG to generate the required amplitude modulation. **Figure S5** depicts the low-power experimental setup. The Mini-Circuits amplifiers provided ~32 mW (15 dBm) power. The LO and AFG frequencies were finely adjusted to minimize reflections. Signal amplitudes were carefully controlled to remain within the oscilloscope's dynamic range (below 5V). For comparison, we also applied a monochromatic step pulse excitation, and the resulting FWD and REF signals were recorded and saved for post-processing. These experiments confirmed the effectiveness of the VCC mechanism in reducing reflections by 70%, as described in the Results section of the main text.

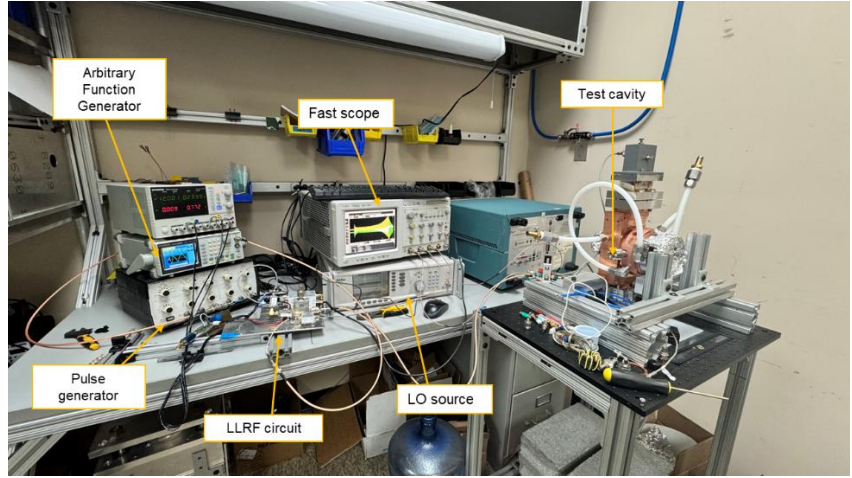

**Figure S5.** Experimental setup for VCC low-power test. This setup included the AFG, pulse generator, Mini-Circuits amplifiers, and a high-speed oscilloscope for signal acquisition.

**High-Power Experimental Validation**

Following the successful demonstration of the VCC regime at low power, we proceeded to high-power testing. The high-power experiment was conducted in a radiation-shielded bunker at RadiaBeam, where the resonator was conditioned to operate under vacuum at elevated power levels. The same methodology used in the low-power experiment was applied here: we first simulated the FWD and REF signals, then calculated the reflection coefficient to assess VCC performance.

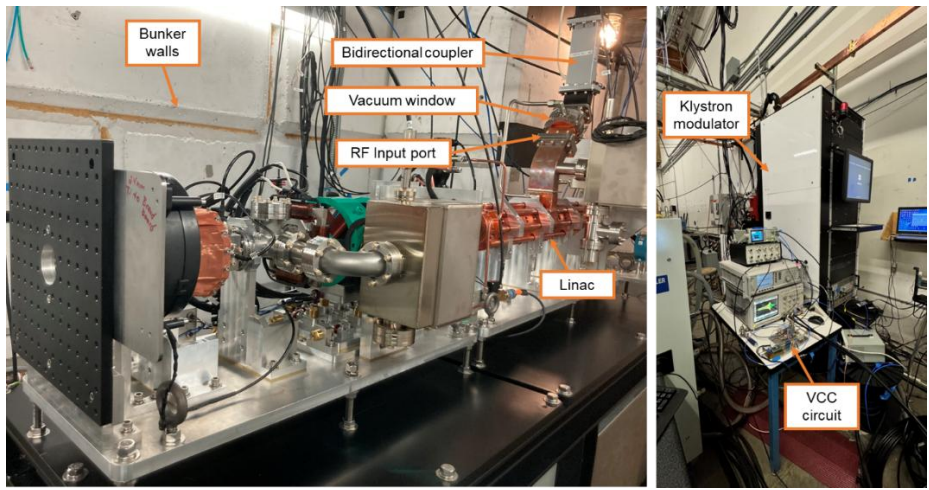

**Figure S6.** Experimental setup for VCC high-power test. The left panel shows the linac housed inside the radiation-shielded bunker, while the right panel displays the LLRF circuit connected to the 5 MW klystron amplifier.

**Figure S6** presents the high-power experimental setup, including the 5 MW ScandiNova K1 klystron amplifier and the linac. During the VCC test, the input power was incrementally increased to 600 kW, delivered in 4 µs pulses at a repetition rate of 1 Hz. To enable oscilloscope measurement of the high-power signals, two -80 dB attenuators were placed at the outputs of the bidirectional coupler, ensuring signal amplitudes remained within the oscilloscope's measurable range. As discussed in the main text, the high-power results demonstrated a ninefold reduction in reflections, confirming the scalability of the VCC mechanism for practical applications in accelerator and fusion technologies.